\documentclass[hyper]{JHEP3}

\input{epsf}
\usepackage{epsfig}
\usepackage{amssymb}
\usepackage{amsfonts}
\usepackage{amsbsy}

\def\IC{\mathbb{C}}

\def\IZ{{\mathbb{Z}}}
\def\IR{{\mathbb{R}}}
\def\IP{\mathbb{P}}

\def\CN {{\cal N}}

\def\CD {{\cal D}}

\def\CP {{\cal P }}

\def\CW {{\cal W}}

\renewcommand{\Im}{{\rm Im }}

\def\one{{\hbox{ 1\kern-.8mm l}}}

\def\To{\rightarrow}
\def\vp{\varphi}

\def\Im{{\rm Im}}

\title{Wall crossing in local Calabi Yau manifolds}

\author{Daniel~L.~Jafferis and Gregory~W.~Moore\\
 NHETC and Department of Physics and Astronomy,
Rutgers University,\\
Piscataway, NJ 08855--0849, USA\\
\\
{\tt jafferis@physics.rutgers.edu, gmoore@physics.rutgers.edu} }

\abstract{We study the BPS states  of a $D6$-brane wrapping the
conifold and bound to collections of $D2$ and $D0$ branes. We find
that in addition to the complexified K\"ahler parameter of the rigid
$\IP^1$ it is necessary to introduce an extra real parameter to
describe BPS partition functions and marginal stability walls. The
supergravity approach to BPS state-counting gives a simple
derivation of results of Szendr\H{o}i concerning Donaldson-Thomas
theory on the noncommutative conifold. This example also illustrates
some interesting limitations on the supergravity approach to BPS
state-counting and wall-crossing.  }

\begin{document}

\begin{section}{Introduction}

One of the most effective tools in the trade of  counting   BPS
states of D-branes wrapped on Calabi-Yau manifolds is the approach
using multi-centered solutions of the low energy effective
supergravity \cite{Denef:2000nb, Denef:2000ar}. This approach leads
to simple and intuitive derivations of wall-crossing formulae for
BPS indices \cite{DM}. There has also been a great deal of activity
on the closely related subject of wall-crossing formulae for
generalized Donaldson-Thomas invariants in the mathematical
literature. In this paper we use the supergravity formulae to make
contact with one set of mathematical results, due to B. Szendr\H{o}i
\cite{Szendroi,BenYoung}. A simple variation of our arguments should
reproduce the related results of  J. Bryan and B. Young
\cite{Young:2008hn}. In the process we highlight some interesting
ways in which the supergravity techniques can break down.

Here is a brief overview of the paper:  In section 2  we   review
the definition of the D6/D2/D0 partition function both from the
microscopic viewpoint, and from the macroscopic - or supergravity -
viewpoint. In section 3 we summarize  previous calculations of
Donaldson-Thomas theory on the conifold at large K\"ahler class as
well as on a noncommutative resolution of the conifold. In section 4
we describe a method for applying wall-crossing techniques to local
Calabi-Yau manifolds. Curiously, we find it necessary to use an
extension of the complexified K\"ahler moduli space of the local
Calabi-Yau  by an extra real parameter $\varphi$. In section 5 we
determine the walls of marginal stability for D6/D2/D0 bound states
on the resolved conifold, and compute the D6/D2/D0 partition
function for certain ranges of $\varphi$.

As a steadfast reader will discover, this paper raises several
unanswered questions. Perhaps the most interesting of these are to
be found in section 6. There we explore some puzzles associated with
crossing a wall of threshold stability.

During the final stage of preparation of this work, we received
\cite{NagaoNakajima,Nagao} in which a similar partition function of
D6/D2/D0 bound states on the resolved conifold is computed using
different techniques.

\end{section}

\begin{section}{Review of BPS black holes and wall crossing}

Consider IIA string theory compactified on a Calabi-Yau 3-fold $X$.
The resulting effective supergravity in $3+1$ dimensions has $\CN =
2$ supersymmetry. There are $(h_{11}(X) + 1)$ independent $U(1)$
vector fields    obtained from the Kaluza-Klein reduction of the RR
fields along even harmonic forms. There exist half-BPS black holes
in these theories, which are the supergravity description of BPS
bound states of D-branes. Roughly speaking these correspond to
solutions of Hermitian Yang-Mills equations on holomorphic cycles in
$X$. Such supersymmetric boundary conditions are thought to be
equivalent to the   ``stable'' objects in the bounded derived
category of $X$. They are characterized by their charges under the
$U(1)$ vector fields. For the purposes of this paper these charges
can be thought of as elements of $H^{even}(X;\IR)$; we adhere to the
conventions of \cite{DM}.

In addition to the ordinary single centered extremal charged black
holes, there exist multi-centered BPS solutions with mutually
nonlocal constituents  \cite{Denef:2000nb,Denef:2000ar}. These
configurations are (generically) true bound states: the binding
energy is negative. Moreover, once the overall center of mass is
factored out,  the classical moduli space of such supergravity
solutions is compact.

In the present work, we will be concerned with a special case of
such objects, namely those with a single unit of D6 charge. It is
convenient to define the generating function of the index of BPS
states with such charges as a formal power series:
\begin{equation}\label{eq:Zd6d2d0}
 Z(u,v; t_\infty)
:= \sum_{N \in \IZ, \beta \in H^4(X,\IZ)} u^N v^\beta \Omega(1-\beta
+ N dV; t_\infty).\end{equation}
Here $dV$ is a generator of $H^6(X;\IZ)$ and   we have turned on
chemical potentials $v$ and $u$ for the D2 and D0 charges,
respectively. We denote the complexified K\"ahler modulus by
$t=B+iJ$, and $t_\infty$ refers to the boundary conditions at
spatial infinity for these moduli.  Note that by a shift of
$B$-field we can, without loss of generality, assume that the $D4$
charge is zero.

\subsection{Index of BPS states in $\CN = 2$ supergravity}

The index of   BPS states $\Omega(1-\beta + N dV; t_\infty)$
appearing in (\ref{eq:Zd6d2d0})   is a piecewise constant integer
function of the asymptotic  moduli $t_\infty$. Moreover,  $\Omega$
can only jump when the asymptotics of the effective potential on the
configuration space change. The only known way that this can occur
is when the physical size of a multi-centered solution diverges for
some values of the asymptotic K\"ahler moduli.

The multi-centered BPS black hole solutions found by Denef  (see
\cite{DM,DdBvdB} for notation) are encoded by a harmonic function
$H: \IR^3 \rightarrow H^{even}(X, \IR)$   with poles at the centers
of the constituents. The metric is given by
\begin{equation}
ds^2 = - e^{2U} (d\tau+\omega)^2 + e^{-2U} d{\vec x}^2,
\end{equation}
where the warp factor (and the spatially-dependent Calabi-Yau
moduli) are determined by the attractor equation
\cite{Ferrara:1995ih,Strominger:1996kf}:
\begin{equation}\label{eq:localattr}
2 e^{-U} \Im(e^{-i\alpha} \Omega_{\rm nrm} ) = -H.
\end{equation}
Here $\alpha$ is the argument of the central charge of the total (RR
gauge theory) charge of the bound state at the background values of
the K\"ahler moduli, $\alpha({\vec x}) := \arg\left( \sum_i
Z(\Gamma_i; t ({\vec x})) \right)$. The normalized periods are given
at large volume by $\Omega_{\rm nrm} = -\sqrt{\frac{3}{4
J^3_\infty}} e^{B+iJ}$. The one-form $\omega$ is determined by
 \begin{equation}\ast_3 d \omega = \langle dH, H \rangle\end{equation}
 where the Hodge star is defined in flat $\IR^3$, and  we use the
symplectic form on $H^{even}(X,\IR)$ given by mirror symmetry (or,
equivalently, by the natural symplectic form on $K^0(X)$.)

 For a configuration with $n$ centers of charge $\Gamma_i$,
the harmonic function is given by
\begin{equation}\label{eq:Hcenters}
H({\vec x}) = \sum_{i=1}^n \frac{\Gamma_i}{|{\vec x} - {\vec x_i}|}
- 2 \Im(e^{-i\alpha} \Omega_{\rm nrm})|_{r=\infty}. \end{equation}
 The asymptotic
value of $H(\vec x)$ for $\vec x \to \infty$ will be denoted by
$H_\infty$. It is related by the attractor equations to the
asymptotic values of the K\"ahler moduli, $t_\infty$.

For a two centered solution of this type, the distance between the
centers can be calculated to be \cite{Denef}
\begin{equation}\label{radius}r_{12} = \frac{\langle \Gamma_1,
\Gamma_2 \rangle}{2 \Im(e^{-i\alpha_\infty} Z(\Gamma_1;
t_\infty))}.\end{equation} This quantity must be positive for the
solution to exist - the resulting condition on $t_\infty$ is called
the Denef stability condition. Furthermore, at a wall of marginal
stability in the K\"ahler moduli space, where $Z(\Gamma_1; t_\infty)
= \gamma Z(\Gamma_2; t_\infty),$ with  $\gamma \in \IR_+$, the
radius $r_{12}$ diverges. The bound state thus decays as $t_\infty$
crosses such a real codimension one wall.

At a wall of marginal stability corresponding to a
decay $\Gamma \to \Gamma_1 + \Gamma_2$, the BPS index will have a
discrete jump given by \cite{DM}
\begin{equation} \Delta \Omega (\Gamma, t) = (-1)^{\langle\Gamma_1,
\Gamma_2\rangle-1} \ |\langle\Gamma_1,\Gamma_2\rangle | \
\Omega(\Gamma_1,t_{\rm ms})\Omega(\Gamma_2, t_{\rm ms}),
\end{equation} when $\Gamma_1$ and $\Gamma_2$ are primitive, and $t
= t_{\rm ms}$ is the point (assumed generic) where the wall is
crossed. In this paper we will also need a generalization to decays
where one of the constituent charges is primitive, but the other is
not. This semi-primitive wall crossing formula is also given in
\cite{DM},
\begin{equation}\label{semiprim} \Omega(\Gamma_1;t) + \sum_{N\geq 1} \Delta \Omega(\Gamma_1 + N \Gamma_2; t)
q^{N} = \Omega(\Gamma_1; t) \prod_{k\geq
1}(1-(-1)^{k\langle\Gamma_1, \Gamma_2\rangle }q^k)^ {k
|\langle\Gamma_1,\Gamma_2\rangle | \Omega(k \Gamma_2; t)}.
\end{equation}
In this paper we will not need the more elaborate
Kontsevich-Soibelman wall-crossing formula
\cite{KSToAppear,Gaiotto:2008cd}.

It is worth noting that the regions of Denef stability are bounded
by walls of marginal stability as well as by walls of \emph{anti}
marginal-stability, where
$$\arg\left(Z(\Gamma_1; t_\infty)\right) =
- \arg\left(Z(\Gamma_2; t_\infty)\right).$$ Note that in general,
the Denef stability condition for a two centered solution,
$$\langle \Gamma_1, \Gamma_2 \rangle {\rm Im}\left(Z^*(\Gamma_2;
t_\infty) Z(\Gamma_1; t_\infty)\right) > 0,$$ is invariant under
$(\Gamma_1, \Gamma_2) \rightarrow (\Gamma_1, - \Gamma_2)$. Clearly
the marginal-stability walls for $\Gamma_1$ with $\Gamma_2$ are
anti marginal-stability walls for $\Gamma_1$ with $-\Gamma_2$, and
{\it vice versa}.

It is natural to conjecture that at any given value of the K\"ahler
moduli, it is impossible for both $(\Gamma_1, \Gamma_2)$ and
$(\Gamma_1,-\Gamma_2)$ to exist as two centered BPS bound states.
This can be established near the boundaries of the region of Denef
stability, since  it is impossible for a two centered solution with
charges $\Gamma_1$ and $\Gamma_2$ to exist near the wall of
anti-marginal stability (and likewise the bound state $\Gamma_1$ and
$-\Gamma_2$ cannot exist near the marginal stability wall of
$\Gamma_1$ with $\Gamma_2$). Assuming that such a solution did
exist, the separation between the centers (\ref{radius}) would
diverge as $t_\infty$ approached the anti marginal-stability wall.
However the total energy of the bound state is given by
\begin{equation}\label{eq:WrongIneq}
\left| Z(\Gamma_1 + \Gamma_2; t_\infty)\right| \rightarrow \left| \
| Z(\Gamma_1; t_\infty)| - | Z(\Gamma_2; t_\infty) |\ \right| <  |
Z(\Gamma_1; t_\infty)| + | Z(\Gamma_2; t_\infty) | , \end{equation}
so energy conservation would be violated in the decay of this state.

Moreover, when  $|Z(\Gamma_1; t_\infty)| \gg |Z(\Gamma_2;
t_\infty)|$ we have
 $\alpha_\infty =
\arg(Z(\Gamma_1+\Gamma_2; t_\infty) \approx \arg(Z(\Gamma_1;
t_\infty)$, even away from the walls.  It then follows that if we
replace $(\Gamma_1, \ \Gamma_2) \rightarrow (\Gamma_1, \ -
\Gamma_2)$, there will be little change in $\alpha_\infty$, yet the
``fragment'' $\Gamma_2$ has been replaced by its anti-particle. We
have just argued that such a two-body solution cannot exist near the
marginal stability wall for $(\Gamma_1,\Gamma_2)$. Therefore, either
something goes wrong with the supergravity solution based on
(\ref{eq:Hcenters}), or there are new walls which ``censor'' such
solutions from existing near the  marginal-stability wall of
$(\Gamma_1,\Gamma_2)$,  or there are interesting re-arrangements of
the boundstate configurations as $t_\infty$ moves towards the
anti-marginal wall. (In this third possibility what we have in mind
is that the boundstate should change - without a change of index -
to   a boundstate of $(\Gamma_1 -(n+1)\Gamma_2)$ and $n \Gamma_2$
for some positive $n$.) It would be interesting to understand better
what really happens.
 \footnote{We thank F. Denef for helpful remarks suggesting ways in
 which the supergravity solution could become singular.  }

\end{section}

\begin{section}{Review of calculations of D6/D2/D0 partition
functions by toric techniques}

\subsection{``Large radius'' Donaldson-Thomas theory on toric Calabi-Yau 3-folds}

The microscopic worldvolume theory of a D6 is the $6+1$ dimensional
DBI action. When the brane is wrapped on a curved Calabi-Yau
manifold, the supersymmetric index can be calculated in the
topologically twisted version of the low energy limit of this
theory. This is a $6$ dimensional Euclidean topological theory
obtained by  twisting the ${\cal N} = 2$ supersymmetric $U(1)$
Yang-Mills theory in 6 dimensions.

There are singular instantons in this topological gauge theory which
carry nonvanishing D2 charge $[F \wedge F] \in H^4(X;\IZ)$, and D0
charge $[F \wedge F \wedge F]\in H^6(X;\IZ)$, where $F$ is the field
strength of the D6 worldvolume abelian gauge field. These singular
instantons are studied in \cite{qfoam}, and are shown to correspond
to ideal sheaves on the Calabi-Yau manifold. (See, however, the
comments in section 4.2 below.)

In \cite{qfoam,MNOP} the Donaldson-Thomas partition function on
the resolved conifold was computed using equivariant localization
on the moduli space of ideal sheaves, building on the work of
\cite{AKMV,topcrystal}. The result is
\begin{equation}\label{DT}
 Z_{\rm DT}(u,v) = M(-u)^2 \prod_{j>0} (1-(-u)^j v)^j,\end{equation}
  where the McMahon function is defined by
  \begin{equation}\label{eq:McMahon}
  M(-u) := \prod_{j>0} \frac{1}{(1-(-u)^j)^j}.
  \end{equation}
This is a special case of the relation between the Donaldson-Thomas
partition function and the topological string partition function on
any Calabi-Yau 3-fold, expressed in terms of the BPS invariants
(a.k.a. the genus zero Gopakumar-Vafa invariants), \begin{equation}
Z_{\rm DT} = \prod_{\beta_h, n_h} \left(1 - (-u)^{n_h} v^{\beta_h}
\right)^{n_h n^0_{\beta_h}}.\end{equation}

We shall see that this is the partition function of D6/D2/D0 bound
states only in a certain limit of the asymptotic K\"ahler moduli. As
in \cite{DM}, this limit requires a specific tuning of the B-field,
even at large volume. We will see more examples of the need for such
tuning below.

\subsection{Szendr\H{o}i's calculation of DT invariants on the noncommutative conifold}

The partition function of Donaldson-Thomas theory on a
noncommutative deformation of the conifold was studied in
\cite{Szendroi}. By definition, the ideal sheaves on this
noncommuative space are cyclic representations of the
noncommutative conifold algebra, $${\cal A} =\mathbb{C}[f_0,f_1]
\langle A_1,A_2,B_1,B_2 \rangle / \langle B_1A_iB_2-B_2A_iB_1,
A_1B_iA_2-A_2B_iA_1, i=1,2 \rangle,
$$ which is a resolution of the singular conifold algebra,
$\textrm{Spec}(\mathbb{C}[x_1,x_2,x_3,x_4]/(x_1x_2-x_3x_4))$. The
meaning of the angular brackets is that products of letters $A_i$
and $B_j$ which do not form paths in the quiver, such as $A_1
A_2$, are set to zero in the algebra. The elements $f_0$ and $f_1$
are idempotent, with $f_0 f_1 = 0$, and are associated to length
zero paths based at the two nodes.


There is again a torus action on the moduli space of ideal
sheaves, which was exploited to reduce the calculation of the
Euler character to the fixed points. These were in one to one
correspondence with pyramid partitions in a length $1$ empty room
configuration. Using combinatoric methods, the generating function
was determined to be \cite{Szendroi, BenYoung} $$Z_{\rm Sz} =
M(-u)^2 \prod_{j>0} (1-(-u)^j v)^j (1-(-u)^j v^{-1})^j.$$

We shall see later that this is exactly the partition function of
D6/D2/D0 BPS states in a particular chamber of the K\"ahler moduli
space.   Certainly a nonzero $B$-field is related to
noncommutative gauge theory
\cite{Schomerus:1999ug,Seiberg-Witten}, so it is not unexpected
that turning on a (``large'') $B$-field should produce this
result. It would be worthwhile understanding in more detail why
the particular value we find is the appropriate one.


\end{section}

\begin{section}{Extended K\"ahler moduli space for local Calabi-Yau}

In  this section we motivate our extension of the K\"{a}hler moduli
space of the local geometry by a single real parameter. Physically,
this parameter measures the strength of a component of
the B-field normal to the local $\IP^1$.


To make sense of the brane charges in the local setting, which is
crucial for being able to apply wall crossing formulae, we will
consider embedding the local geometry into a compact Calabi-Yau. We
proceed to determine the behavior of the central charges of various
D-branes in the local limit, in which, informally, all K\"ahler
parameters are taken to be large, with the exception of the size of
the rigid curve under investigation. There is a considerable
simplification of the dependence of the index of BPS D6/D2/D0 states
on the K\"ahler moduli in that limit, analogous to that found in the
large volume limit \cite{DM}. We will find that,   in addition to
the complexified K\"ahler parameter of the $\IP^1$, an extra real
parameter remains in the formulae for marginal stability walls and
BPS indices, even after we take the decompactification limit.

To  motivate further the extension of the K\"ahler moduli space of
the local curve, we will show  that it emerges naturally from the
worldvolume perspective as well.  This worldvolume theory can be
made well-defined in the case of noncompact Calabi-Yau manifolds by
using techniques of toric localization in the fiber. We shall see
that this leads, in a somewhat different way, to the same picture of
an extension of the local K\"{a}hler moduli space by an additional
real parameter.

\subsection{Motivation from taking a limit of compact CY}

Our main example will be the resolved conifold. However, let us
begin with a more general setting.  We consider the local limit of a
compact Calabi-Yau 3-fold, $X$, in which the only homology class
which remains small is a rigid rational curve, dual to $\beta \in
H^4(X)$. The K\"{a}hler parameter is
\begin{equation}\label{Kahler} t = z \CP + \Lambda e^{i\vp}
\CP', \end{equation}
 where $\CP \beta = 1$, $\CP' \beta = 0$, and $\Lambda$ is a positive real number.
 We
are interested in the behavior in the $\Lambda \To +\infty$ limit.
We assume that the positive class $\CP \in H^2(X;\IR)$ and
semi-positive class $\CP' \in H^2(X;\IR)$ are such that $t$ is in
the K\"ahler cone for all positive $\Lambda$. We also assume
$(\CP')^3
>0$ (and hence $\CP'^2 \neq 0$). \footnote{  This excludes the
possibility that the surface dual to $\CP'$ is a K3 fiber. } Note
that in these variables $t$ lies in the K\"ahler cone for $\vp \in
(0, \pi)$ and $\Im (z)
>0$.

The central charges in the local limit are easily obtained from the
large volume expression for the periods. Consider a charge
$$\Gamma_1 = 1 - m' \beta + n' dV.$$ Then its central charge
is given in the local limit by $$Z(\Gamma_1; t) = \Lambda^3
e^{3i\vp} -m' z -n' \longrightarrow \Lambda^3 e^{3i\vp}.$$

 The multi-centered solutions which can contribute to the index of
states with charge $\Gamma_1$ have a core $\Gamma_{core} = 1$,
together with a ``halo'' of particles or ``fragments'' of charge
$\Gamma_h = -m_h \beta + n_h dV$. The latter have central charges
given by
$$Z(\Gamma_h; t) = -m_h z - n_h.$$

Therefore we see that the position of the walls of marginal
stability for charges of the above form only depends on
the coordinates $z \in \IC$ and $\vp \in (0, \pi)$. 
The extra data needed to define the index in our noncompact setting
is the variable $\varphi$, which encodes, in the local limit, all of
the dependence of the central charges on the K\"ahler moduli
introduced in the compactification. By definition, the $B$-field is
given by $B ={\rm Re}(z) \CP + \Lambda \cos\vp \CP'$. In the
noncompact limit, it is more meaningful to talk about the local
density of the $B$-field along those directions normal to the local
curve, normalized with respect to the local value of the K\"ahler
form. This is exactly $\cot \vp$.


\subsection{Motivation from worldvolume instanton equation}

We have seen by embedding the local geometries we wish to study into
a compact Calabi-Yau manifold that the D6/D2/D0 partition function
depends on an additional parameter, $\vp$, which comes from the
K\"{a}hler structure in the noncompact directions. It would be
gratifying to understand how this occurs from the worldvolume point
of view. This will also lead us to an intrinsic definition of the
D6/D2/D0 partition function as a function of the background moduli
in a local Calabi-Yau 3-fold, without the need for a global
completion.

Note that the central charge of the D6 brane cannot be calculated
directly in the local geometry, due to its divergent volume; this
is the reason we regulated by compactifying above. However, the
argument of the central charge does have a meaning in this
context, and this is all one needs to find the walls of marginal
stability. To see this, we go back to the original worldvolume
gauge theory description of D6/D2/D0 bound states.

Following the work of \cite{qfoam} we expect that for large K\"ahler
class  our bound states can be described as singular instantons of
the $U(1)$ twisted ${\cal N}=2$ gauge theory living on the 6-brane
worldvolume. They are given by singular solutions of the hermitian
Yang-Mills equations,
\begin{equation}\label{eq:u1instanton}
F^{2,0} = 0, \ \ \ \ F^{1,1}\wedge J^2 = \ell J^3,
\end{equation}
which can be made well-defined by turning on a noncommutative
deformation. In \cite{qfoam} it was argued that counting solutions
to (\ref{eq:u1instanton}) is equivalent to counting  ideal sheaves
and hence amounts to enumeration of standard Donaldson-Thomas
invariants.

However, there is a problem with this picture: It does not account
for walls of marginal stability extending to infinity. \footnote{A
similar remark applies to the partition function counting D4/D2/D0
boundstates on Calabi-Yau manifolds with $h^{1,1}(X)>1$
\cite{Diaconescu:2007bf,DdBvdB,Andriyash:2008it}.} Thus the picture
can be at best correct in certain chambers of the complexified
K\"ahler cone. (These chambers will be located in  Section 5.) One
might suspect that one should apply - on the algebro-geometric side
- some stability conditions,    generalizing the DUY theorem that
slope-stable holomorphic vector bundles are in correspondence with
solutions to the Hermitian Yang-Mills equations \cite{UY,Donaldson}.
But slope stability  is trivial for ideal sheaves, and
correspondingly, the solutions to (\ref{eq:u1instanton}) do not
depend on background moduli.

Of course, we should in fact be applying $\Pi$ stability (a.k.a.
Bridgeland stability) to elements of the   derived category of
coherent sheaves of charge  $1-m\beta+n dV$
\cite{Douglas:2000ah}.\footnote{An alternative stability condition
is that studied in \cite{Pandharipande:2007qu}.} There should be a
generalization of the DUY theorem in which stable objects in the
derived category (at large K\"ahler class) are in 1-1 correspondence
with solutions of some differential equations generalizing the
Hermitian Yang-Mills equation. The natural expectation from physics
is that those should be (at large K\"ahler class)  the full
non-linear instanton equations, with non-vanishing $B$-field, found
in \cite{MMMS}. (We may still neglect nonperturbative worldsheet
instanton corrections, since these are subleading in the local
limit.)  The equations from \cite{MMMS} can be written in the form
\begin{equation}\label{eqMMMS} {\rm Re}\left[ e^{F + t} \sqrt{Td(X)}\right]^{(6)} = \cot 3
\vp \ \Im \left[ e^{F + t} \sqrt{Td(X)}\right]^{(6)},\end{equation}
where we take the 6-form components, and recall that  $t = B + i J$.
In a compact Calabi-Yau manifold, this equation would be integrated
to determine the number $\cot (3\vp)$.  For local Calabi-Yau
3-folds, this integration is not well-defined, and $\vp$ becomes a
free parameter. We expect that the  Euler character of the moduli
space of solutions depends on $\vp$. Given the formula for the
central charge, if we compactified the local geometry in some way,
$3\vp$ would become the argument of the central charge of the
6-brane, so this is the same $\vp$ we introduced above. It would be
very interesting to confirm that the torus equivariant calculation
of the Euler character of the moduli space of such instantons in the
noncompact resolved conifold agrees with the index of BPS states
obtained from the local limit
of a compact Calabi-Yau manifold. 

Note that the solutions to the non-linear equations are also
singular, for the trivial reason that a smooth configuration with no
4-brane charge must satisfy $ch_1 = 0$ so that $F = d A$ if $F$ is
smooth, and hence $[F \wedge F]$ and $[ F\wedge F\wedge F]$ must
also vanish. 
However, the D2 and D0 charges carried by this singular solution are
finite (which is not the case for the linearized equations)
\cite{Seiberg-Witten}. Thus it might be possible to analyze the
moduli space of instantons for the non-linear equations without
making a noncommutative deformation.

In principle, one should be able to see explicitly   the
appearance and disappearance of solutions (with  such mild
singularities) of equation (\ref{eqMMMS}) as one varies the local
K\"{a}hler modulus, $z$ as well as $\vp$. We have not attempted to
do this so we content ourselves with finding the walls where the
central charges are aligned, and using the supergravity inspired
wall crossing formulae of \cite{DM}. This will turn out to
reproduce exactly the answer computed at the 
conifold point by counting equivariant quiver representations by
\cite{Szendroi,BenYoung,Young:2008hn}.

\end{section}

\begin{section}{Walls and their crossing}

\subsection{Which are the relevant walls?}

In the local geometry, only the pure D6 brane exists as a single
centered object, its attractor flow hitting a zero at the conifold
point. Note that using the large volume limit of the periods, this
attractor flow would hit the boundary of the K\"ahler moduli space
at the point $z=0$, but the periods receive significant instanton
corrections in that regime, modifying the result. The existence of
this BPS state is established by microscopic reasoning, as is the
index of D2/D0 particles. The single centered attractor flows for
other charges of the form $1-m\beta+n dV$ always end on a zero at
a regular point in the moduli space. The fact that there are no
other single centered solutions carrying one unit of D6 charge
(and potentially various D2 and D0 charges) is due to the
vanishing of all higher genus Gopakumar-Vafa invariants in the
resolved conifold.

The partition function of D6/D2/D0 bound states will receive
contributions from a variety of single and multi-centered bound
states. Our strategy for computing the index as a function of the
background K\"ahler moduli will be to start with a known result, and
determine the partition function in other chambers using wall
crossing formulae. In particular, the higher genus Gopakumar-Vafa
invariants vanish in the resolved conifold, hence there is a unique
single centered object carrying one unit of D6 charge, namely the
pure D6 brane itself. We will begin our analysis in the core region
of the moduli space, where all multi-centered solutions are
unstable, and $Z_{D6D2D0} = 1$.

First we will determine which split attractor flows can decay in
the region of moduli space we are interested in. The local limit
imposes strong constraints on the central charges, and we will
find that only a particular class of fragments can become
marginally stable. This will lead to a halo picture of the
D6/D2/D0 bound states, similar to the description of the same
system at large volume on one parameter Calabi-Yau manifolds in
\cite{DM}.

Suppose there is a split attractor flow for $\Gamma_1 + \Gamma_2 =
\Gamma = 1 - m \beta + n dV$. Let $\Gamma_1 = a + \CD - \beta_h +
n_h dV$, where $a \in H^0(X;\IR)$, $\CD \in H^2(X;\IR)$, $\beta_h
\cdot \CP=m_h$ and $\beta_h \cdot \CP' = M_h$. The wall of marginal
stability for this flow in the local limit is located where the
central charges
$$Z_1 = \frac{a}{6} \Lambda^3 e^{3i\vp} + \frac{\CD}{2} \cdot
(\Lambda e^{i\vp} \CP'+z \CP)^2 - M_h \Lambda e^{i\vp} - m_h z -
n_h,$$ and $$Z_2 = \frac{1}{6} \Lambda^3 e^{3i\vp} - m z - n -Z_1,$$
are aligned. In the $\Lambda \To \infty$ limit, these expressions
simplify, and we keep only the leading terms.

It is  obvious that no walls with $a > 1$ (or $a<0$) can extend into
the local regime, since then the D6 charges of the two fragments
would have opposite sign, and $Z_1$ would be anti-aligned with
$Z_2$, those being the dominant terms. Hence the quantization of
charge implies that either $a=0$ or $a=1$. At the expense of
reversing the roles of $\Gamma_1$ and $\Gamma_2$, we can assume that
$a=0$.

Generically, if $\CD \neq 0$ then $\CD \cdot \CP'^2$ will be
non-vanishing, and in the local limit, $Z_1 = \frac{1}{2} \CD \cdot
\CP'^2 \Lambda^2 e^{2i\vp}$, and $Z_2 = \frac{1}{6} \Lambda^3
e^{3i\vp}$. These are never aligned inside the K\"ahler cone.
Moreover,   by assumption,  $\CP'$ is semipositive and $\CP'^2$ is
non-vanishing, so $\CP'^2$ is dual to an element of the Mori cone
(i.e. a positive sum of effective curve classes). Thus if $\CP'^2
\CD=0$ then $\CD$ is not very ample, and cannot support a single
centered bound state. Since in this limit we have already shown that
$D6$-$\overline{D6}$ bound states are unstable, it appears that such
fragments do not exist in the regime of interest. Thus we conclude
that $\CD=0$.


Furthermore, for $M_h \neq 0$, the associated term would be the
dominant one, and the wall of marginal stability would be located at
$e^{2i\vp} \ || \ -M_h$, which extends into the K\"ahler cone in the
local limit only along the line $\vp = \pi/2$ (if $M_h>0$). We will
see later that our entire analysis, beginning in the core region,
can be applied for $\vp < \pi/2$, so we will not cross these walls
if they are present.

The physical picture that emerges from the above arguments is that
in the limit of the background K\"ahler moduli that we are
considering, the attraction between D6 branes and all other charges,
with the exception of $-m_h \beta + n_h dV$, is very large, and any
such BPS multi-centered solutions have radii that decreases with a
power of $\Lambda$. Thus only D2/D0 fragments may become unbound as
the K\"ahler moduli are dialed, with a resulting jump in the index
of BPS states.

Therefore, all of the splits in the attractor flow tree involve a
charge of the form $\Gamma_h = -m_h \beta + n_h dV$. Moreover, as we
argued earlier, the only single centered configuration with D6
charge is the pure D6,\footnote{This is true for the resolved
conifold geometry itself, due to the absence of high genus GV
invariants. Any compact Calabi-Yau manifold in which the local curve
is embedded will have a nontrivial spectrum of core states. However
our argument above shows that there are no jumps in the index of
those core states, so we can consider the compactification
independent ratio of the D6/D2/D0 partition function in the various
chambers we identify in the local limit to this $Z_{core}$. This is
what we calculate, and identify as the partition function on the
resolved conifold.} so it must be one of the ends of the split
attractor tree. The BPS mass of the pure D6 is much greater than
that of the D2/D0 fragments, thus the bound states we are
considering look like ``atoms''  with D2/D0 particles in halos of
various radii around the pure D6 core. This is the special case of
the picture found in \cite{DM} in a similar context, in which only a
single core state exists in the spectrum.

We have shown that the only walls of marginal stability that extend
into the region of large $\Lambda$ are those for the charges $1 - m'
\beta + n' dV$ and $-m_h \beta + n_h d V$. Given the expressions for
their central charge, the location of such walls only depends on the
parameters $z$ and $\vp$ in the $\Lambda \To \infty$ limit.
Therefore, assuming the conjecture that the index of BPS bound
states can only change across walls of marginal stability, we can
conclude that the index of BPS D6/D2/D0 bound states has a well
defined limit, depending only on $z$ and $\vp$:\footnote{It is
important that the local curve is rigid, otherwise the degeneracies
of the D2/D0 fragments themselves (the Gopakumar-Vafa invariants) in
the compact Calabi-Yau manifold, $X$, would differ from their values
in the equivariant regulation of the noncompact local Calabi-Yau
manifold. In the case of the resolution of local $A_1 \times \IC$,
this difficulty can be overcome by partially compactifying to toric
local rigid surface, such as $\IP^1 \times \IP^1$, and then
embedding in a compact Calabi-Yau manifold.}

\begin{equation}\label{degen} \lim_{\Lambda \rightarrow \infty}
\Omega_X(1-m \beta + n dV; t) := \Omega_{\rm local}(1 - m \beta + n
dV; z, \vp).\end{equation}

\subsection{Location of the walls of marginal stability}

We wish to determine the location of walls of marginal stability for
$\Gamma_1 = 1 - m'\beta +n' dV$ and $\Gamma_2 = -m\beta+n dV$ in the
effective K\"ahler moduli space we discussed in section 4. Using the
above formulae for the central charges, we see that they are aligned
when $$\vp = \frac{1}{3} \arg (-m z -n) + \frac{2\pi}{3} k,$$ for
$k\in \IZ$. These are to be thought of as real codimension 1 walls
in the three real dimensional space defined by $z$ and $\vp$. Note
that the walls are independent of the charges $n'$ and $m'$ in the
local limit, as their contribution is subleading to that of the D6.
 There will be different cases for the four possible signs of $m$
and $n$. For future reference, we will denote the walls of marginal
stability as
\begin{equation}
\CW^m_n = \{(z,\vp): \vp = \frac{1}{3} \arg(z + n/m) +
\frac{\pi}{3}\}
\end{equation}
\begin{equation}
 \CW^{-m}_n = \{(z,\vp): \vp = \frac{1}{3} \arg(z - n/m)\}
 \end{equation}
 \begin{equation} \tilde\CW^{-m}_n = \{(z,\vp):\vp = \frac{1}{3} \arg(z-n/m) +
\frac{2\pi}{3}\}, \end{equation}
where $m > 0$.  The
structure of the walls turns out to be most clearly visible by
fixing $z$ and varying $\vp$ in its allowed range from $0$ to $\pi$.

To connect this three dimensional moduli space with the one K\"ahler
parameter analysis of \cite{DM}, it is useful to consider a special
class of K\"ahler forms given by $t= x J_0$, where $x \in \IC$ and
$J_0$ is real. Such geometries have $B$ proportional to $J$ as
vectors in $H^2(X;\IR)$, and the periods naturally correspond to
those found in the one parameter case. In our variables, this
condition implies that $\vp = \arg(z)$.

Therefore we can identify the core region, which is the chamber
containing the limit of large volume and zero $B$-field, where in
our case only the pure D6 exists (due to the absence of higher genus
GV invariants in the resolved conifold), as an open region around
$\vp = \pi/2$, $z = i y$ for sufficiently large positive $y$,
following the results of \cite{DM}. There are no relevant walls
between this point and $\vp = \frac{1}{3} \arg(z) + \frac{\pi}{3}$
(for any value of $z$ in the upper half plane) , so we shall start
our analysis there, where $Z_{D6/D2/D0} = 1$.

\FIGURE{ \centerline{
\includegraphics[height=4cm]{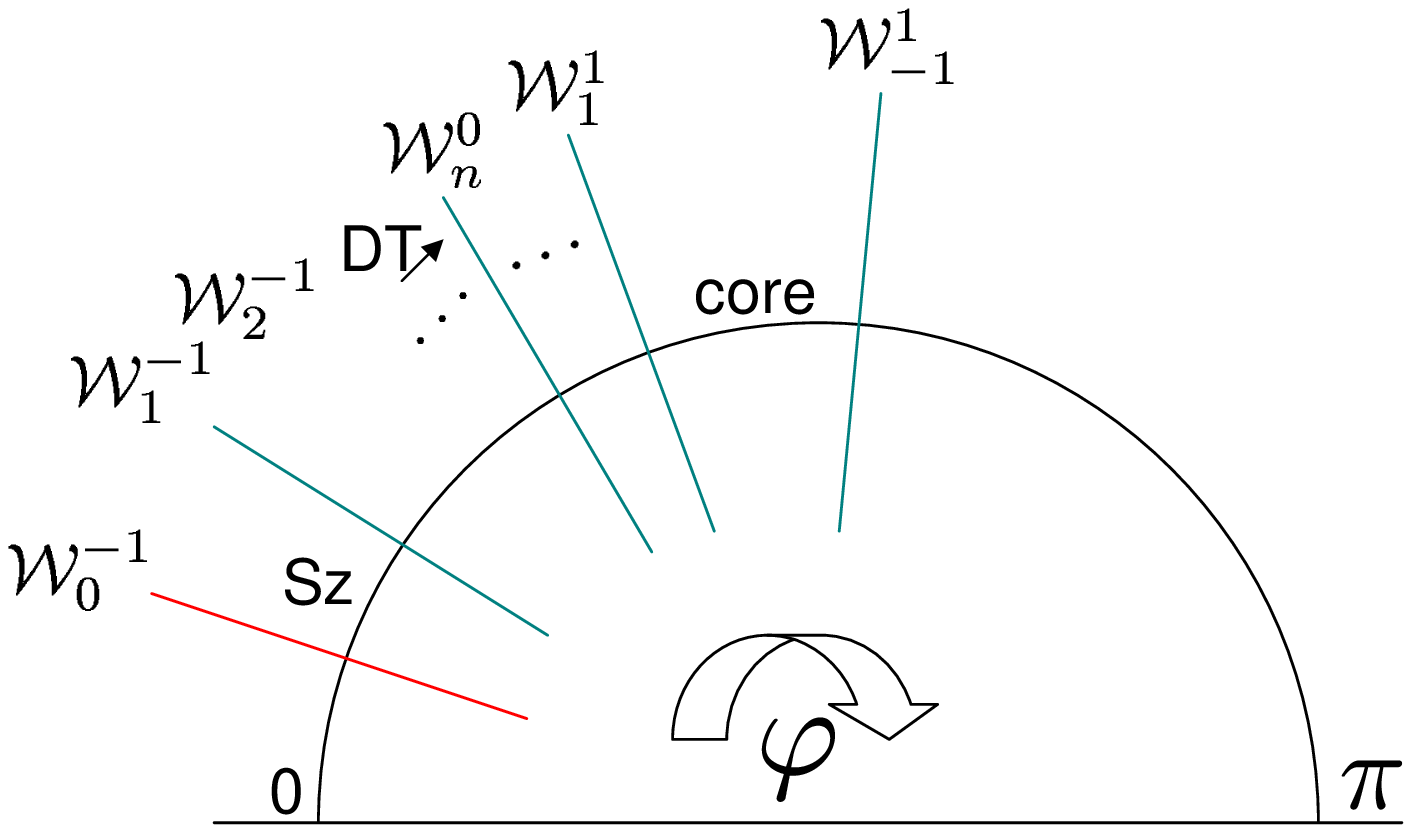}} \caption{This
figure illustrates the position of the walls of marginal stability
at fixed $z$. The core region, Donaldson-Thomas limit, and
Szendr\H{o}i region are labelled.} \label{Walls} }


\subsection{Computing the partition function in all the chambers}

In the resolved conifold, the BPS indices are known, and the
degeneracies of primitive D2/D0 particles can be easily extracted.
In particular, the only non-vanishing ones are
\cite{Gopakumar:1998ii,Gopakumar:1998jq}
\begin{equation}
\Omega(\pm \beta + n dV) = 1
\end{equation}
valid for all $n$, and
\begin{equation}
\Omega( n dV) = - 2,
\end{equation}
valid for  $n \neq 0$,
 and these degeneracies have no walls of
marginal stability at large radius by arguments similar to those
used in section 5.1. The D2/D0 fragments give a net fermionic
contribution to the index, and their contribution to the D6/D2/D0
partition function will be identical to that of particles obeying
Fermi-Dirac statistics, while the index of the D0 fragments is net
bosonic, so their contribution to the partition function will be
identical to that of particles obeying Bose-Einstein statistics.

Thus we only need to consider the walls $\CW^{\pm 1}_n$, $\CW^0_n$,
which are simple enough to draw; see figure \ref{Walls}. The
chambers between successive walls will be denoted as
$[\CW^{m_1}_{n_1} \CW^{m_2}_{n_2}]$ where we assume
$\CW^{m_1}_{n_1}$ is to the left of $\CW^{m_2}_{n_2}$ in the $\vp$
direction.  Starting in the core region, as we decrease $\vp$,
increasing the $B$-field in the directions normal to the local
curve, the collection of walls for $\Gamma_1$ with $-m\beta+n dV$
with $n, m>0$ are crossed first. They are located at $\vp =
\frac{1}{3} \arg(z - |n/m|) + \pi/3$. Note that $0 < \frac{1}{3}
\arg(z + |n/m|) < \frac{1}{3} \arg(z - |n/m|) < \pi/3$, so these are
indeed the first walls we cross. A necessary condition for the bound
state $\Gamma_1+\Gamma_2$ to exist is the
 Denef stability criterion:
\begin{equation}
\langle \Gamma_1, \Gamma_2 \rangle \Im (Z_1 Z_2^*) = - n \Im
(e^{3i\vp} (-m z^* - n))
> 0.
\end{equation}
Clearly this is satisfied when   $\vp$ is less than $\frac{1}{3}
\arg(z - |n/m|) + \pi/3$. A closer examination of the supergravity
solution reveals no pathologies and we conclude that  the bound
state enters the spectrum as $\varphi$ crosses the wall in the
direction of smaller values. There are no nontrivial core states
in this local model. Therefore, in the chamber $[{\cal W}^1_{n+1}
{\cal W}^1_n]$, the partition function is   given by
\begin{equation} Z^+_n(u,v) = \prod_{j=1}^n (1-(-u)^j v)^j ,
\end{equation}
in accord with the semiprimitive wall crossing formula (\ref{semiprim}).

There are an infinite number of walls $\CW^1_n$, one for each $n >
0$, however the index for any given total charge remains finite as
$n \to \infty$, since   the successive fragments have ever
increasing D0 charge, and eventually do not contribute to the index
for a fixed total charge. These walls have an accumulation point at
$\vp = \pi/3$, which is the location of the wall of marginal
stability of D6 with pure D0. The limit from the right,
\begin{equation}
\lim_{n \To \infty} Z_n^+(u,v) =\prod_{k>0} (1-(-u)^k v)^k = Z_{\rm
DT}'(u,v) \end{equation}
 is the reduced Donaldson-Thomas partition
function defined in \cite{MNOP}.

 Continuing to increase the $B$-field in the normal directions, the
 central charges of the D6 bound state aligns with that of
 $\overline{D2}/D0$ fragments when $\vp = \frac{1}{3} \arg(z - |n/m|)$ for $m<0$,
 $n>0$. In the chamber $[{\cal W}^{-1}_{n} {\cal W}^{-1}_{n+1}]$, from the semiprimitive wall crossing formula
 (\ref{semiprim}) we
 have the D6/D2/D0 generating function \begin{equation} Z_n^-(u,v)
 = M(-u)^2 \prod_{k>0} (1-(-u)^k v)^k \prod_{m > n} (1-(-u)^m
 v^{-1})^m .\end{equation}
The factor of $M(-u)^2$ appears when crossing the D6D0 wall at $\vp
= \pi/3$. Note that the walls $\CW^{-1}_n$ also accumulate at $\vp =
\pi/3$ for $n\to \infty$,  so that the  limit from the left is
\begin{equation}
\lim_{n\to \infty} Z_n^-(u,v) = M(-u)^2 \prod_{k>0} (1-(-u)^k v)^k
= Z_{\rm DT}(u,v) , \end{equation}
where $Z_{\rm DT}$ is given in equation (\ref{DT}).

Continuing to cross all the walls $\CW^{-1}_n$ to smaller values of
$\vp$   we enter the region, $\frac{1}{3} \arg z < \vp < \frac{1}{3}
\arg (z-1)$ where the partition function is given
 by the Szendr\H{o}i form,
\begin{equation}\label{eq:SzPF}
Z_{\rm Sz} = M(-u)^2 \prod_{k
> 0} (1 - (-u)^k v)^k (1- (-u)^k v^{-1})^k .
\end{equation}
 We will refer to this as the Szendr\H{o}i region.

The conifold point
 lies in the boundary of the Szendr\H{o}i region as $\Im(z)$ goes to zero.
 Note that (\ref{eq:SzPF})
 is precisely the partition function computed by Szendr\H{o}i by
 counting quiver representations at the conifold point.
 Continuing to $\vp < \frac{1}{3} \arg z$ leads to a breakdown in
 the supergravity approximation, and results in puzzles whose
resolution is beyond the scope of this paper. We will discuss
these puzzles in the next section.

It is natural to speculate that an analogous behavior would be
observed in the D6/D2/D0 on other local curves. In particular, the
analysis of the positions of the walls of marginal stability in
the effective three real dimensional moduli space we defined did
not depend  on the triple intersection numbers of the local
Calabi-Yau manifold. Thus we are motivated to conjecture that in
the analog of the Szendr\H{o}i region (some neighborhood on one
side of the codimension one wall $\vp = \frac{1}{3} \arg z$), the
D6/D2/D0 partition function will have the form
\begin{equation} Z(u,v) = M(-u)^\chi \prod_{\beta_h, n_h} \left(1
- (-u)^{n_h} v^{\beta_h} \right)^{n_h n^0_{\beta_h}} \left(1 -
(-u)^{n_h} v^{-\beta_h} \right)^{n_h n^0_{\beta_h}},
\end{equation} in terms of the Gopakumar-Vafa invariants. This
structure has been observed in local $A_1$ in \cite{Young:2008hn}
using techniques similar to \cite{Szendroi} at the orbifold point.

\end{section}

\begin{section}{The D6-D2 threshold stability wall, and some
puzzles}

In this section we discuss some puzzles that arise when we consider
the supergravity approach to wall-crossing in the vicinity of $\vp =
\frac{1}{3} \arg z$.  On this wall the central charges of the D6 and
$\overline{D2}$ branes become aligned. Nevertheless,  the index is
not expected to change since these charges are mutually local:
$\langle 1, \beta \rangle = 0$. Physically, there are no bound
states of a D6 and D2, so there is no halo whose radius diverges.
This locus of alignment of the central charges of mutually local BPS
states is called a ``threshold stability wall'' \cite{DdBvdB}. (See
also \cite{Aspinwall:2006yk} for a discussion of such walls.)

One might consider continuing the analysis of the partition function
across the walls ${\cal W}^{-1}_{-n}$ at $\vp = \frac{1}{3} \arg (z
+ n)$ for $n > 0$. These walls have an accumulation point at $\vp =
0$. When decreasing $\vp$, one would be passing from the region of
Denef stability to a region of instability. This immediately
presents a problem, since no bound states with net negative D0
charge are present in the spectrum we found in the Szendr\H{o}i
region.


Now, recall the discussion centered on (\ref{eq:WrongIneq}). It
follows from those remarks that  there cannot be any halo
solutions containing $-\beta+ n dV$ fragments for $n>0$ orbiting a
D6 core as the walls $\CW^{-1}_{-n}$ are approached. This suggests
that the halo picture of the bound states must break down in the
region around $\vp = \frac{1}{3} \arg
 z$. We shall see that indeed a number of surprising phenomena occur at
 precisely this value of $\vp$.

First, we note that the halo radii associated to all of the $\pm
\beta + n dV$ and $n dV$ fragments become equal at this threshold
wall, suggesting a potential mechanism for a reorganization of the
spectrum contributing to the index. This will be found to be
reflected in the degeneration of the  split attractor flow trees for
our D6/D2/D0 bound states as $t_\infty$ approaches the wall. Next,
all of the D6/D2/D0 solutions, including the pure D6 brane solution,
will be shown to exit the regime of validity of supergravity at the
threshold wall. The reason for this is that   the spatially
dependent K\"ahler modulus of the $\IP^1$ crosses the boundary of
the K\"ahler cone somewhere in the interior of the solution.



Applying (\ref{eq:Hcenters}) we see that the supergravity solution
describing a D6 brane surrounded by D2/D0 particles in halos is
described by
\begin{equation} H(\vec{x}) = H_\infty + \frac{1}{|\vec{x}|} +
\sum_i \frac{-\beta_i + n_i dV}{|\vec{x}-\vec{x_i}|},
\end{equation} where the halo radii are determined from the
integrability conditions to be
\begin{equation}\label{radii}
|\vec{x_i}| = r_i = \frac{-n_i}{\Im \left(e^{-3i\vp} (m_i z + n_i)
\right)} \sqrt{\frac{ J_\infty^3}{3}}.
\end{equation}
This simple result is valid because all of the
fragments are mutually local, and hence produce no forces on each
other. The positivity of the above radii is exactly the Denef
stability condition. Note that this does not distinguish whether a
given D2/D0 particle or its anti-particle (related by $n_i, \ m_i
\rightarrow -n_i,\ -m_i$) will be present in a supersymmetric bound
state. Let $\rho$ denote the radius of the pure D0 halos. These are
obtained by putting $m_i=0$ to give
\begin{equation}
\rho = \frac{1}{\sin 3\vp} \sqrt{ J_\infty^3\over 3}.
\end{equation}
In terms of $\rho$ we can write
\begin{equation}\label{eq:rirho}
r_i = \rho \frac{\sin 3\vp}{\sin 3\vp + \frac{m_h}{n_h}
\Im(e^{3i\vp} z^*)}.
\end{equation}
At the threshold stability wall of D6 and $\overline{D2}$ we see
that the radii of all halos coincide, since by (\ref{eq:rirho}) $r_i
= \rho $ when $\arg z = 3 \vp$. Different components of the moduli
space of supergravity solutions with   equal total charge become
connected at this location. For example, consider the bound state,
$I$, of $\Gamma_{core}=1$, $\Gamma_1 = -\beta+dV$, and $\Gamma_2 = +
\beta + dV$, and another bound state, $II$, of $\Gamma_{core}=1$ and
$\Gamma_3 = 2 dV$. These  contribute to the same index. In the
Szendr\H{o}i region, for $\vp > \frac{1}{3} \arg z$, the fragments
orbit in halos of different radii that scale like
$\Lambda^{\frac{3}{2}}$, so these two multi-centered solutions are
disconnected. At the threshold wall, the radii are equal, and the
subspace of the moduli space of  configurations of type $I$ where
the fragments $\Gamma_1$ and $\Gamma_2$ have the same angular
position is indistinguishable from the moduli space of
configurations of type $II$.

A similar degeneration can be observed in the split attractor flow
trees for our D6/D2/D0 bound states. Consider the split attractor
flow for a total charge $\Gamma = 1 - m\beta + n dV$ into
$\Gamma_h = -m_h \beta + n_h dV$ and $\Gamma - \Gamma_h$. The
attractor flow of $\Gamma$, beginning at $t_\infty$, is described
by the attractor flow equation
\begin{equation} 2 e^{-U} {\rm Im} (e^{-i\alpha} \Omega_{\rm nrm}) = 2 {\rm Im}(e^{-i\alpha_\infty}
\Omega_{\rm nrm}|_\infty) - \Gamma \tau,\end{equation} where the
parameter $\tau$ is related to the spatial position by $\tau = 1 /
r$. The branching point of the tree is located along the flow at
$\tau = \frac{1}{r_h}$, where the split occurs. At the threshold
wall, we saw that (for $\Lambda \to \infty$) all $r_h$ are equal,
independent of $m_h$ and $n_h$, thus the branch points for any such
D2/D0 fragment are coincident. But by definition, the split is
located at the intersection of the attractor flow for $\Gamma$ with
the wall of marginal stability for $\Gamma$ and $\Gamma_h$.

Thus when $t_\infty$ is located on the D6-$\overline{D2}$ wall, the
branch point of the split attractor flow tree is located at the
intersection of the walls of marginal stability of $\Gamma$ with all
charges of the form $-m_h \beta + n_h dV$. However it is impossible
for the central charges $Z(-m_h \beta + n_h dV; t) = -m_h z - n_h$
for be aligned for all values of $m_h$ and $n_h$. Therefore the
branch point of the split flow must be hitting the zero of the exact
function  $Z(\Gamma; t)$. In general, this zero is located at small
volume, which suggests that the supergravity approximation is
breaking down, assuming that the graph of the split attractor flow
in the K\"ahler moduli space is a subset of the full range of moduli
in the associated supergravity solution.


To  establish fully the existence of a supergravity solution, one
must check that the local K\"ahler moduli remain in the K\"ahler
cone, and that the local discriminant is everywhere positive. We can
parameterize the harmonic function governing the supergravity
solution in the form
\begin{equation} H(\vec{x}) = r e^S \left( 1 - Y^2 + N
dV\right),\end{equation} where $S \in H^2(X;\IR)$, while $Y \in
H^2(Y;\IR)$ is in the K\"ahler cone, and $N \in \IR$.

The standard attractor solution formulae
\cite{Shmakova:1996nz,Moore:1998pn,Denef:2003,DdBvdB} allow us to
express the local moduli in terms of the components of the harmonic
function $H(\vec x)$. We begin by solving
\begin{equation}D_{ABC}\label{ys}
y^A y^B = - 2 H_C H^0 + D_{ABC} H^A H^B,\end{equation}
where   $D_{ABC}$ are the triple intersection numbers for $y^A$ in
the K\"ahler cone. Then we form $Q^3 = \left(\frac{1}{3} D_{ABC} y^A
y^B y^C\right)^2$ and
$$\Sigma = \sqrt{\frac{Q^3-L^2}{(H^0)^2}},$$ where
  $L = H_0
(H^0)^2 + \frac{1}{3} D_{ABC} H^A H^B H^C - H^A H_A H^0$. Finally,
we have
\begin{equation}\Im(t^A) = y^A
\frac{\Sigma}{Q^{\frac{3}{2}}}.\end{equation}

Consider the behavior of the K\"ahler moduli at the radius of the D0
halo, $|\vec{x}| = \rho$.   For any value of the background moduli,
$H^0$ vanishes along the sphere with this radius. Therefore
(\ref{ys}) implies that
\begin{equation}\label{eq:constya}
y^A =  - H^A =  - \sqrt{\frac{3}{J^3_\infty} } \Im(e^{-3i \vp} (z
\CP^A + \Lambda e^{i\vp} (\CP')^A)),
\end{equation}
 where we use the fact that no D4 charge is
present, so the $H^A$ are spatially constant. Here we have chosen
the branch of solutions with the minus sign so that the coefficient
of $\CP'$ is positive, thus $\Im(t^A)
> 0$ for $\varphi > \frac{1}{3} \arg (z)$.

The asymptotic values of the moduli were chosen to be in the local
limit, with the $\IP^1$ small relative to the total volume of the
Calabi-Yau manifold. Given that $\CP' \beta = 0$, we can chose a
positive basis of $H^2(X;\IR)$ such that $(\CP')^1 = 0$ and $\CP^1 =
1$. In that basis the K\"ahler parameter of the $\IP^1$ is exactly
$\Im(t^1)$. On the sphere $|\vec{x}| = \rho$, we then have that the
local value of that modulus is
\begin{equation} \Im(t^1) =
\sqrt{\frac{3}{J^3_\infty}} \frac{\Sigma(\vec
x)}{Q^{\frac{3}{2}}(\vec x)} Im(e^{-3i\vp} z).\end{equation}

In the limit that $H^0 \rightarrow 0$, the discriminant remains
finite, and is given by
\begin{equation}
\Sigma = \sqrt{2 H_A H^A - \frac{H_0}{3} D_{ABC} H^A H^B H^C}.
\end{equation}
On the other hand, $Q$ is nonvanishing, as follows from
(\ref{eq:constya}).  Putting everything together, we find that at
the threshold wall, $\Im(t^1)$ actually vanishes on the sphere of
radius $\rho$, even in the pure D6 solution. This radius is large,
scaling as $\Lambda^{3/2}$ in the local limit, so we are witnessing
a clear breakdown of the supergravity approximation - the K\"ahler
moduli exit the K\"ahler cone. It seems plausible that if one used
the quantum corrected periods, the Calabi-Yau would be in the
flopped conifold phase for $|\vec x| < \rho$. By continuity, as one
approaches the wall $3\vp = \arg z$ from the right, the modulus of
the $\IP^1$ begins to shrink in the interior of the solution.

It is possible that the spectrum of BPS states remains unchanged
across the threshold wall, however the tools of this paper are
insufficient to determine whether that is the case. Let us assume
the validity of the conjecture that the index can only jump at walls
of marginal stability. Suppose that one knew that partition function
in the region $R = [\CW^{-1}_{-2} \CW^{-1}_{-1}]$ as a formal power
series  in $u, u^{-1}$ and $v, v^{-1}$. Then an application of the
semiprimitive wall crossing formula would imply that
\begin{equation}\label{eq:beyondsz}
Z_{\rm Sz} (u,v) = Z(u,v; R) \left(1-(-u)^{-1} v^{-1} \right),
\end{equation}
where we recall that the bound states of $1-m'\beta + n' dV$ with
$+\beta - dV$ are Denef-stable in Szendr\H{o}i region, and
Denef-unstable in the new region, $R$. Moreover   we have used
$\Omega(+\beta - dV) = 1$.

If we formally solve for $Z(u,v;R)$ by multiplying by the inverse
power series $(1+u^{-1}v^{-1})^{-1}$ we obtain  the generating
function
\begin{equation}
Z(u,v; R) = u v M(-u)^2 \prod_{p>1} \left(1-(-u)^p v \right)^p
\prod_{k>0} \left(1 - (-u)^k v^{-1}\right)^k.
\end{equation}
This expression is rather unsatisfactory  since it implies that the
index of BPS states with total charge given by the pure D6 is
vanishing in the chamber $R$. The microscopic theory of the pure D6
is the topologically twisted $U(1)$ gauge theory in six dimensions,
which would seem to have a unique ground state in the sector with no
instantons for any value of the K\"ahler moduli.


One should be wary of such formal manipulations of wall-crossing
formulae. What the formula really states is that  the indices of BPS
states of the two sides of the wall are related by
\begin{equation}\label{Rwallcrossing} \Omega(1-m\beta+n dV; Sz) = \Omega(1-m\beta+n dV;
R) + \Omega(1-(m+1)\beta + (n+1) dV; R).\end{equation}
 It is clear
that, given the partition function in the Szendr\H{o}i region, the
index in the region $R$ is not uniquely determined by this wall
crossing formula. This should be contrasted with the situation when
one knows the BPS indices on the unstable side of a marginal
stability wall, in which case the wall crossing formulae do uniquely
determine the answer on the stable side. The non-uniqueness in our
case can be parametrized by noting that given arbitrary $\Omega(1+ n
dV; R)$, a solution to the wall crossing formula
(\ref{Rwallcrossing}) can be found, with the degeneracies
$\Omega(1-m\beta + n dV; R)$ for $m\neq 0$ now determined.

Indeed, if we insist   that $\Omega(1; R) = 1$ then we find
\begin{eqnarray}
\Omega(1+n\beta-n dV;R) & =& (-1)^n\textrm{ for }n\geq 0,\\
\Omega(1-\beta+dV; R) & = & 0,\\
\Omega(1-n\beta +n dV;R) & = &  (-1)^n\textrm{ for }n\geq 2,
\end{eqnarray}
%
%
%
%
%
by solving the equations (\ref{Rwallcrossing}) recursively. This
also seems problematic, at least in the supergravity picture, since
the BPS index is now non-vanishing for bound states with exactly
opposite fragment charges. (Recall the discussion at the end of
section 2.) On the other hand, since we have exited the regime of
validity of the supergravity approximation, the counting of BPS
states in this region is beyond the scope of the tools of this
paper. Once again, it would be extremely interesting to understand
what really happens in this regime.

\begin{section}* {Acknowledgements}

We would like to thank Wu-yen Chuang, Frederick Denef, Emanuel
Diaconescu, Davide Gaiotto, and Bal\'{a}zs Szendr\H{o}i for useful
discussions. We would especially like to thank Mina Aganagic. She
asked a good question which initiated this project, and she
participated in the first stages of the project.  This work is
supported by the DOE under grant DE-FG02-96ER40949.

\end{section}

\end{section}

\end{document}